\documentclass[journal]{IEEEtran}

\usepackage{myStyleIEEE}
\usepackage{nomencl}
\usepackage{ragged2e}
\usepackage[thinc]{esdiff}
\IEEEoverridecommandlockouts

\newlength{\bracewidth}

\usepackage{pgf} 
\definecolor{high}{HTML}{FF2800}  
\definecolor{low}{HTML}{ffffff}  
\newcommand*{\opacity}{70}
\newcommand*{\minval}{0.0}
\newcommand*{\maxval}{77.82}
\newcommand{\gradient}[1]{
    \ifdimcomp{#1pt}{>}{\maxval pt}{#1}{
        \ifdimcomp{#1pt}{<}{\minval pt}{#1}{
            \pgfmathparse{int(round(100*(#1/(\maxval-\minval))-(\minval*(100/(\maxval-\minval)))))}
            \xdef\tempa{\pgfmathresult}
            \cellcolor{high!\tempa!low!\opacity} #1
    }}
}

\usepackage{etoolbox}
\renewcommand\nomgroup[1]{%
  \item[\bfseries
  \ifstrequal{#1}{P}{A. Parameters}{%
  \ifstrequal{#1}{V}{C. Variables}{%
  \ifstrequal{#1}{S}{B. Sets and Indices}{}}}%
]}

\begin{document}

\title{Stability Constrained Optimization in High IBR-Penetrated Power Systems{\textemdash}Part II: Constraint Validation and Applications}

\newtheorem{proposition}{Proposition}
\renewcommand{\theenumi}{\alph{enumi}}

\author{Zhongda~Chu,~\IEEEmembership{Member,~IEEE,} and
        Fei~Teng,~\IEEEmembership{Senior Member,~IEEE} 
        
        
\vspace{-0.5cm}}
\maketitle
\IEEEpeerreviewmaketitle

\begin{abstract}
Multiple operational constraints of power system stability are derived analytically and reformulated into Second-Order Cone (SOC) form through a unification method in Part I of this paper. The accuracy and conservativeness of the proposed methods are illustrated in the second part. The validity of the developed constraints is tested against dynamic simulations carried out based on the modified IEEE 39-bus system. Furthermore, the developed power system stability constraints are applied to the optimal system scheduling model. The resulting stability-constrained system scheduling problem aims to achieve the most economical system operation while ensuring different stability in power systems with high Inverter-Based Resources (IBR) penetration. Moreover, based on the stability-constrained optimization model, a novel marginal unit pricing scheme is proposed to quantify the stability services of different units appropriately according to their economic value in maintaining system stability, thus providing rational incentives to the stability service provider and insightful information for the stability market development.
\end{abstract}


\makenomenclature
\renewcommand{\nomname}{List of Symbols}
\mbox{}
\nomenclature[P]{$M$}{system inertia$\,[\mathrm{MWs/Hz}]$}

\section{Introduction}
Power system stability is conventionally analyzed offline on a day-ahead or even longer timescale and carried out only in the worst cases, i.e., peak demand conditions \cite{Li2021}. However, as the IBR penetration increases, this approach may fail due to the highly uncertain nature of renewables. To address this issue, online stability assessment tools have been developed to assess the system stability in near real-time. The authors in \cite{8063445} propose an online rotor angle stability assessment approach based on maximal Lyapunov exponent, which is estimated recursively using a least-squares-based method from real-time Phasor Measurement Unit (PMU) measurements. A robust stability assessment toolbox is developed in \cite{7488976} which is able to provide mathematically rigorous certificates for a multi-machine system while considering the uncertainty in equilibrium points and fault behaviors. 

With the stability assessment tools, instability operation can thus be avoided by generation redispatch as the last line of defense \cite{Li2021}. However, this process is typically implemented by heuristic trial-and-error methods based on engineering experience and judgment, which can be time-consuming and inefficient as the IBR penetration further increases \cite{915388}. As a result, the concept of stability-constrained optimization in power systems has been proposed in literature \cite{9281042} to ensure the power system stability while maximizing the economic benefit. Initially, stability constraints are developed and incorporated into the Optimal Power Flow (OPF) problem, based on system dynamics dominated by conventional Synchronous Generators (SGs) \cite{1216151,867137}. 

System stability constraints are also developed considering the integration of IBRs in more recent research. Frequency stability due to the reduced system inertia and frequency response from SGs are investigated in \cite{9447933,9066910,9201188,9475967}, where the optimal amount of frequency support services are determined by incorporating the post-contingency frequency dynamics into the system scheduling model. 
Voltage stability constraints are derived based on power flow Jacobian matrix singularity in \cite{8279490,9786660} and included into OPF and unit commitment problem respectively. A small-signal stability constrained OPF model in IBR-dominated microgrids is developed in \cite{9511198} where the stability constraint is derived using a Lyapunov stability equation from reduced-order models of microgrids. 

On the other hand, in order to incentivize system devices such as SGs and Grid-ForMing (GFM) IBRs to provide stability services, these units should be properly paid according to their contribution to stabilizing the system. The concept of a stability service market has been proposed in \cite{StabilitMarket}, which is still in progress with the mechanism to price different stability services being unclear. The work in \cite{9855885} proposes a methodology to determine the shadow price of frequency response reserves from IBRs.
A bargaining game-based optimal bidding strategy of virtual power plants in frequency control ancillary services markets is introduced in \cite{9329195}.
However, these works are still limited to the frequency ancillary services market, with other stability services being untouched. 

Since most of the existing works related to stability-constrained optimization focus on single stability, it is not clear how different stability constraints and their interactions influence system operation and the economic performance of the system. Moreover, how to price the stability services when different stability constraints are considered in the optimization model is also unclear. In this context, this paper first validates the system stability constraints developed in Part I based on the modified IEEE 39-bus system and applies them to the system scheduling model, which not only enables the detailed analysis of the impacts of, and interactions among different system stability constraints, but also provides valuable insight to incentivize different grid assets for stability service provision. The main contributions of this paper are summarized below:
\begin{itemize}
    \item The stability constraints developed in Part I of this paper are validated through dynamic simulations in the modified IEEE 39-bus system and the accuracy of the unification process is assessed. 
    \item A technical-economic test system that involves both dynamical models for SGs and IBRs for the stability assessment and the steady-state models for the decision analysis (economic performance) is developed and presented as a benchmark system for future development of linking stability maintenance and economic optimization.
    \item A stability-constrained optimal scheduling framework is formulated by incorporating different stability constraints into the Unit Commitment (UC) model. The interactions among different stability constraints and their impact on the overall system operation cost are investigated for the first time. Computational analysis of different formulations is also performed.
    \item A novel pricing mechanism based on marginal unit price is proposed which for the first time enables the quantification of stability service price of different units, providing valuable insight for the establishment of power system stability market. 
\end{itemize}
    
The rest of the paper is organized as follows. Section~\ref{sec:2} introduces the modified IEEE 39-bus system. The unified stability constraints are validated and assessed in Section~\ref{sec:3}. The applications are described in Section~\ref{sec:4} followed by results in Section~\ref{sec:5}. Finally, Section~\ref{sec:6} concludes the paper. 

\section{Test System Description} \label{sec:2}
\begin{figure}[!b]
    \centering
    \vspace{-0.4cm}
	\scalebox{0.497}{\includegraphics[trim=0 0 0 0,clip]{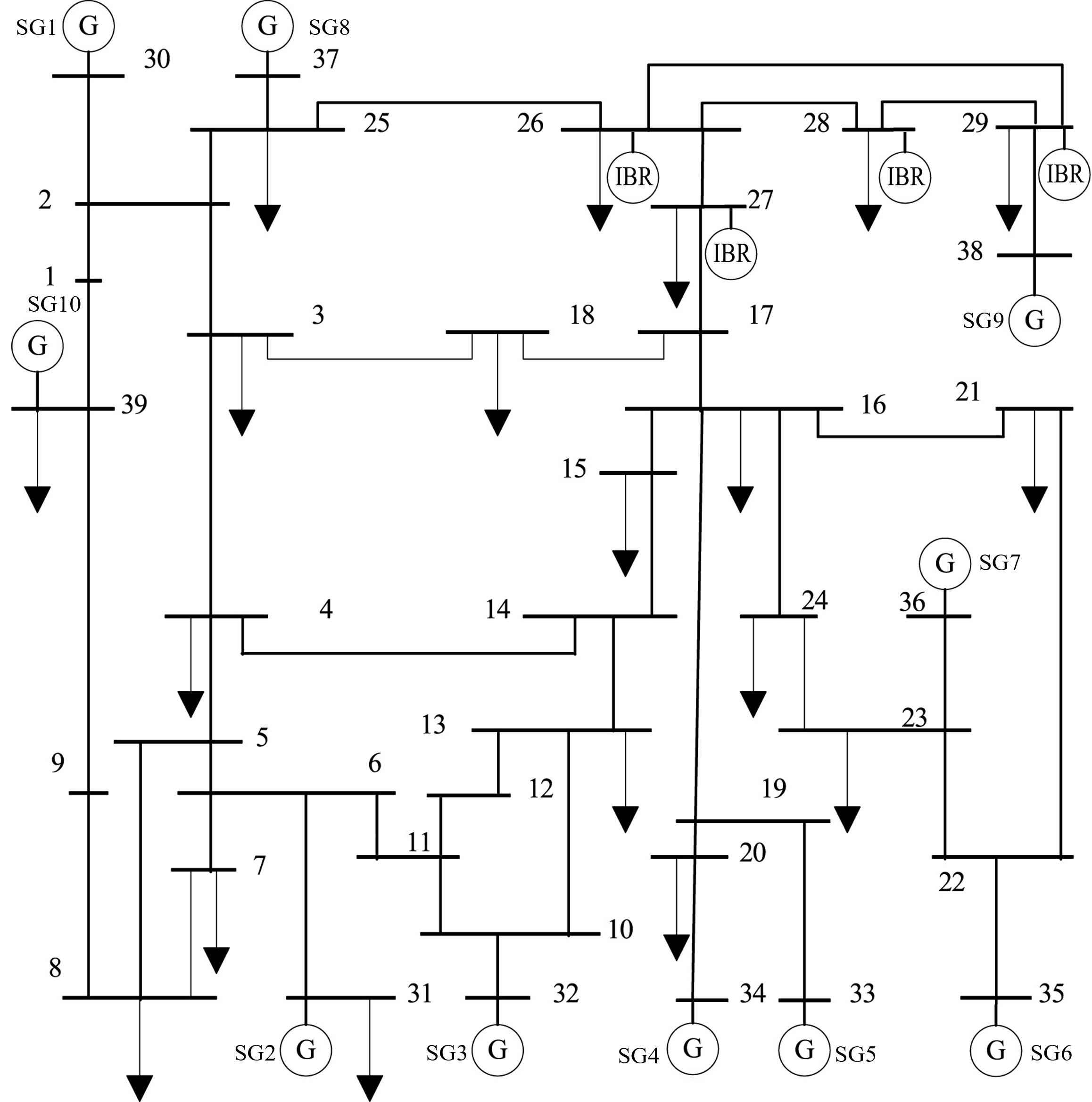}}
    \caption{\label{fig:39-bus}Modified IEEE-39 bus system.}
\end{figure}

The IEEE-39 bus system shown in Fig.~\ref{fig:39-bus} is utilized as the test system for constraints validation and applications. The models and parameters related to the test system are introduced in this section. To increase the renewable penetration, IBRs are added at Bus 26, 27, 28 and 29 with the second one being a GFM unit. Note that we add 4 additional IBRs to the original IEEE 39-bus system, instead of using IBRs to replace the SGs. This is because, on the one hand, we would like to cover a wide range of operating conditions in the case studies. Hence, for the situations where the wind penetration level is low or the IBRs are not allowed to provide any stability services, the system still needs enough SGs to maintain the power balance or the system stability. On the other hand, at higher wind penetration, most of the SGs are not committed online, thus representing an IBR-dominated system. 

For synchronous generators, we consider a traditional model equipped with a prime mover and a \textit{TGOV1} governor. Furthermore, the automatic voltage regulator based on a simplified excitation system \textit{SEXS} is incorporated, together with a \textit{PSS1A} power system stabilizer~\cite{entsoeGen}. Internal machine dynamics are characterized by the transients in the rotor circuits described through flux linkage, as transients in the stator windings decay rapidly and can thus be neglected. The inclusion of stator circuit balance completes the set of differential-algebraic equations, which combined with 6 controller states and swing equation dynamics yields a standard $12^\mathrm{th}$ order system. The main parameters are listed in Table~\ref{tab:paramSM}. For more details regarding the generator modeling and internal parameter computation, we refer the reader to~\cite{Kundur1994}.

\begin{table}[!t]
\renewcommand{\arraystretch}{1.2}
\caption{Generator Control Parameters}
\label{tab:paramSM}
\noindent
\centering
    \begin{minipage}{\linewidth} 
    \renewcommand\footnoterule{\vspace*{-5pt}} 
    \begin{center}
        \begin{tabular}{ c || c | c }
            \toprule
            Parameter & Symbol & Value \\ 
            \cline{1-3}
            Droop control gain & $R_g$ & $5\, \%$\\            
            Governor time constant & $T_g$ & $0.5 \, \text{s}$\\
            Reheat time constant & $T_r$ & $10 \, \text{s}$\\
            Mechanical power gain factor& $K_m$ & $0.85$\\
            Turbine power fraction factor & $F_h$ & $0.1$\\
            Normalized inertia constant & $M_g$ & $6 \, \text{s}$\\
            Normalized damping factor & $D_g$ & $1 \, \text{p.u.}$\\
            Transducer time constant & $T_e$ & $0.05 \, \text{s}$\\
            AVR exciter control gain & $K_{AVR}$ & $200$\\
            Saturation minimum output & $V_f^\text{min}$ & $0$\\
            Saturation maximum output& $V_f^\text{max}$ & $4$\\
            PSS stabilization gain & $K_{PSS}$ & $10$\\
            Washout time constant & $T_{w}$ & $2 \, \text{s}$\\
            $1^\text{st}$ lead-lag derivative time constant & $T_{1}$ & $0.25 \, \text{s}$\\
            $1^\text{st}$ lead-lag delay time constant & $T_{2}$ & $0.03 \, \text{s}$\\
            $2^\text{nd}$ lead-lag derivative time constant & $T_{3}$ & $0.15 \, \text{s}$\\
            $2^\text{nd}$ lead-lag delay time constant & $T_{4}$ & $0.015 \, \text{s}$\\
            \bottomrule
        \end{tabular}
        \end{center}
    \end{minipage}
\end{table}


For IBR dynamics, we consider a state-of-the-art VSC control scheme previously described in \cite{8579100,UrosISGTeurope}, where the outer control loop consists of active and reactive power controllers providing the output voltage angle and magnitude reference by adjusting the predefined setpoints $(x^*)$ according to a measured power imbalance: 
\begin{subequations}
\label{eq:pwr_ctrl}
\begin{align}
    \dot{\omega}_{c} &= \frac{1}{M_{s}}(p^*-p)-\frac{1}{M_{s}}D_{s}(\omega_{c}-\omega^*)  \\
    v_{c} &= v^* + R^q (q^*-\frac{\omega_z}{\omega_z + s}q),
\end{align}
\end{subequations}
with $M_{s}$ and $D_{s}$ being the synthetic inertia and damping, $R^q$ denoting reactive power droop gains, $\omega_z$ representing the LPF cut-off frequency, and $\dot{\theta}_{c}=\omega_{c}\omega_n$. For the GFL control, the above voltage vector reference is replaced by measurements, i.e., $[v^*, \omega^*] = [v_f, \omega_{pll}]$. Subsequently, the reference voltage vector signal $(v_{c}\angle\theta_{c})$ is passed through a virtual impedance block, as well as the inner control loop consisting of cascaded voltage and current controllers. The output is combined with the DC-side voltage in order to generate the modulation signal $m$. In order to detect the system frequency at the connection terminal, a PLL-based synchronization unit is included in the model. However, for the purposes of a grid-forming converter this unit is bypassed via $\omega^*=\omega_n$. The complete mathematical model consists of 13 states for the grid-forming and 15 states for the grid-feeding converter unit, with the inclusion of the LPF current and voltage dynamics. More details on the overall converter control structure and employed parametrization can be found in \cite{8579100,UrosISGTeurope}.


For the static analysis, the parameters of transmission lines, loads and generator capacities are available in \cite{39_bus}. The load and renewable generation profile in \cite{6026941,9968474} is adapted for the simulation during the considered time horizon. The characteristics of thermal generators are given in Table~\ref{tab:SG_para} while considering the data in \cite{9403907,9968474}, with their location being Bus $\{30,37\}$, $\{31,32,33,34,35,36,38\}$ and $\{39\}$ respectively. Other system parameters are set as follows: load demand $P^D\in [5.16, 6.24]\,\mathrm{GW}$, load damping $D = 0.5\% P^D / 1\,\mathrm{Hz}$, base power $S_B = 100\mathrm{MVA}$. The test system for dynamic and economic analysis is publicly available here \cite{Testsystem}. 

\begin{table}[!t]
\renewcommand{\arraystretch}{1.2}
\caption{{Parameters of Thermal Units}}
\label{tab:SG_para}
\noindent
\centering
    \begin{minipage}{\linewidth} 
    \renewcommand\footnoterule{\vspace*{-5pt}} 
    \begin{center}
        \begin{tabular}{ c || c | c | c }
            \toprule
            Type & Type I & Type II & Type III \\ 
            \cline{1-4}
            No-load Cost [k\pounds/h]& 4.5 & 3 & 0\\
            Marginal Cost [\pounds/MWh]& 47 & 200 & 10\\
            Startup Cost [k\pounds]& 10 & 0 & N/A \\
            Startup Time [h]& 4 & 0 & N/A\\
            Min Up Time [h] & 4 & 0 & N/A\\
            Min Down Time [h] & 1 & 0 & N/A \\
           \bottomrule
        \end{tabular}
    \end{center}
    \end{minipage}
\end{table}

\section{Stability Constraint Validation} \label{sec:3}
In Part I of this paper, stability constraints in high-IBR penetrated systems are derived analytically and reformulated into SOC form through a unified SOC reformulation. In this section,  dynamic simulations based on Matlab/Simulink are first carried out to validate the stability constraints, followed by an approximation assessment of the unification process to demonstrate its accuracy and conservativeness. These stability constraints are then visualized in terms of the system generation mix through a simplified example to showcase the relationship and interaction of various constraints.

\subsection{Validation of Stability Criteria}\label{sec:2.1}
\begin{figure}[!b]
    \centering
    \vspace{-0.4cm}
	\scalebox{1.2}{\includegraphics[trim=0 0 0 0,clip]{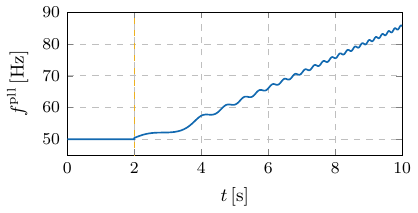}}
    \caption{\label{fig:EMT_EQ}PLL frequency with ($t\le2\,\mathrm{s}$) and without ($t>2\,\mathrm{s}$) equilibrium point.}
\end{figure}

Stability constraints introduced in Part I of this paper are validated here based on time-domain simulation conducted in the modified IEEE-39 bus system. The existence of the equilibrium point of PLL dynamics of GFL IBRs is considered first, with the PLL frequency shown in Fig.~\ref{fig:EMT_EQ}. At $t=2\,\mathrm{s}$, the grid impedance is increased such that the PLL of the GFL IBR at Bus 29 cannot regulate $v_q$ to zero (loss of equilibrium point). The PLL frequency diverges after that, which illustrates that a stable operation of PLL cannot be achieved without the existence of an equilibrium point. 

\begin{figure}[!t]
    \centering
	\scalebox{1.2}{\includegraphics[trim=0 0 0 0,clip]{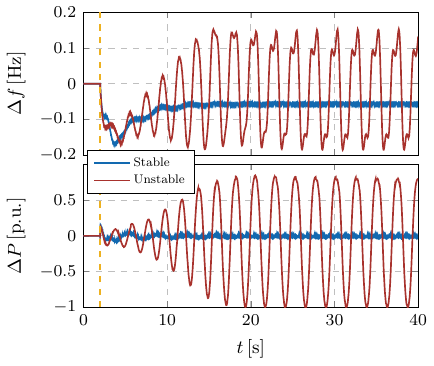}}
    \caption{\label{fig:EMT_SSS}System dynamics after small disturbance at IBR terminal at $t=2\,\mathrm{s}$: (i) frequency deviation; (ii) active power deviation.}
\end{figure}

The generalized short circuit ratio ($\mathrm{gSCR}$) developed in \cite{8488538} is utilized to characterize the small-signal synchronization stability. A small disturbance is applied to the output power of the IBR at Bus 29 at $t=2\,\mathrm{s}$ to trigger the dynamics. The trajectories of IBR frequency deviation $\Delta f$ and active power deviation $\Delta P$ are plotted in Fig.~\ref{fig:EMT_SSS}. In the case of $\lambda_{\mathrm{min}}>\mathrm{gSCR}_{\mathrm{lim}}$, the system converges to a stable operating point after the disturbances, represented by the blue curves whereas in the unstable case $\left(\lambda_{\mathrm{min}}<\mathrm{gSCR}_{\mathrm{lim}}\right)$, both the frequency and active power start to oscillate with an increasing magnitude when subjected to the disturbance. 

\begin{figure}[!t]
    \centering
    \vspace{-0.4cm}
	\scalebox{1.2}{\includegraphics[trim=0 0 0 0,clip]{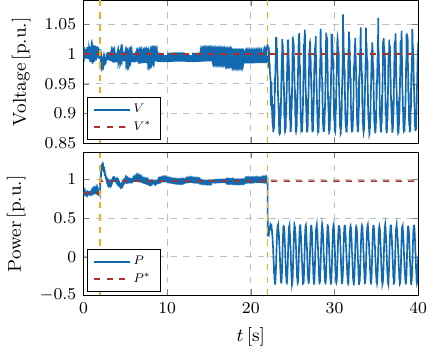}}
    \caption{\label{fig:EMT_VS}System dynamics after step increment of IBR power references: (i) voltage trajectory; (ii) active power trajectory.}
\end{figure}

The static voltage stability of GFL IBRs is characterized by the active power transfer capability and system strength. The associated time domain simulation is demonstrated in Fig.~\ref{fig:EMT_VS} where a step increase of $0.15 \,\mathrm{p.u.}$ is applied to the active power reference $P^*$ of the IBR at Bus 29 at $t=2\,\mathrm{s}$. As indicated by the blue curve in Fig.~\ref{fig:EMT_VS}-(ii), when the system strength is sufficient, the actual active power output could follow the reference and stabilizes fast and the voltage in Fig.~\ref{fig:EMT_VS}-(i) also stays close to the reference. However, at $t=22\,\mathrm{s}$, the system strength is reduced by disconnecting a parallel line between the IBR and the grid. This triggers the voltage instability immediately where the voltage and active power become oscillatory, meaning that the required active power cannot be transferred to the grid with the current system strength.

\begin{figure}[!t]
\captionsetup[subfigure]{labelformat=empty}
  \centering
    \begin{subfigure}{0.485\textwidth}
        \centering
        \scalebox{0.97}{\includegraphics[trim=0 0 0 0.1,clip]{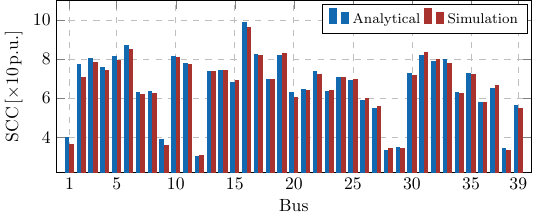}}
        \vspace{-0.35em}  
        \caption{}
        \nonumber
        \label{fig:EMT_SCC}       
        \vspace{-1.9em}
    \end{subfigure} 
    \begin{subfigure}{0.485\textwidth}
        \centering
        \scalebox{0.97}{\includegraphics[trim=0 0 0.3 0,clip]{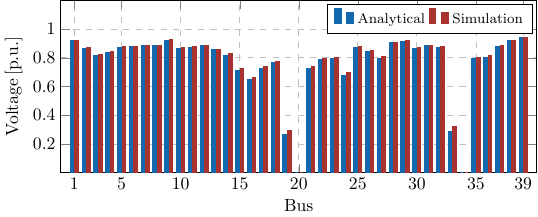}}
        \vspace{-0.35em}  
        \caption{}
        \label{fig:EMT_dV}       
        \vspace{-2em}
    \end{subfigure} 
  \caption{\label{fig:plots}System frequency evolution after a small step disturbance at $t=1\,\mathrm{s}$: (i) system SCC at different buses; (ii) bus voltages after three-phase short circuit fault at Bus 20.}
\end{figure}

Three-phase balanced short circuit fault is applied to different buses in the system in order to validate the short circuit current calculation. The results of dynamic simulation and analytical calculation are depicted in Fig.~\ref{fig:plots}. The SCC at different buses of the system presents a wide distribution in Fig.~\ref{fig:plots}-(i), varying from around $30\,\mathrm{p.u.}$ to $100\,\mathrm{p.u.}$, depending on their electrical distance to SGs. The shortage of SCC is identified to be appearing at the buses near the IBRs and far away from SGs such as Bus 29 and 12. The post fault voltage after a three-phase short circuit fault at Bus 20 is also demonstrated in Fig.~\ref{fig:plots}-(ii), where lower voltages can be spotted at the buses near the fault. Nevertheless, a good approximation of the analytical results can be observed for both SCC and post-fault voltages.

\subsection{Accuracy Assessment of Unification Process} \label{sec:2.2}
Recall that the unification process is essentially the process of determining the optimal SOC parameters $\mathcal{K} =\{\mathsf{A}_i, \mathsf{b}_i, \mathsf{c}_i, \mathsf{d}_i\},\,\forall i$ of $\Tilde{\mathbf{g}}_i$, the estimated expression of the original nonlinear function $\mathbf{g}_i$, which can be defined as below:
\begin{align}
    \label{g_SOC}
    \Tilde{\mathbf{g}}_i  (\mathsf{X}_i) & = \mathsf{c}_i \mathsf{X}_i+ \mathsf{d}_i - \left\lVert \mathsf{A}_i \mathsf{X}_i+ \mathsf{b}_i \right\lVert,
\end{align}
where $\mathsf{X}_i$ is the variables in system operation and matrices $\mathsf{A}_i\in\mathbb{R}^{j\times \dim( \mathsf{X}_i)}$, $\mathsf{b}_i\in\mathbb{R}^{j}$, $\mathsf{c}_i\in\mathbb{R}^{1\times\dim( \mathsf{X}_i)}$, $\mathsf{d}_i\in\mathbb{R}$ are parameters to describe the shape and position of the SOC. The value of $j$ is a degree of freedom that should be chosen to balance the trade-off between complexity and accuracy. To find the most accurate approximation while maintaining the conservativeness, these parameters can be obtained by solving the following optimization problem:
\begin{subequations}
\label{DM3}
\begin{align}
    \label{obj3}
    \min_{\mathcal{K}}\quad & \sum_{\omega^i \in \Omega^i_2} \left(\mathbf{g}_{i}^{(\omega^i)} - \Tilde{\mathbf{g}}_{i}^{(\omega^i)} \right)^2\\
    \label{coef_ctr2}
    \mathrm{s.t.}\quad & \Tilde{\mathbf{g}}_{i}^{(\omega^i)}< \mathbf{g}_{i_{\mathrm{lim}}},\,\,\forall \omega^i \in \Omega^i_1\\
    \label{coef_ctr3}
    &\Tilde{\mathbf{g}}_{i}^{(\omega^i)} \ge {\mathbf{g}}_{i_{\mathrm{lim}}},\,\,\forall \omega^i \in \Omega^i_3,
\end{align}
\end{subequations}
with $\omega^i = \{\mathsf{X}_i^{(\omega^i)},\, \mathbf{g}_{i}^{(\omega^i)}\}\in \Omega^i$ denoting the entire data set. An example of SCC constraint unification accuracy is assessed with the results shown in Table~\ref{tab:SCC_unification}, where two types of error as defined below are analyzed.
\begin{itemize}
    \item Type I: insecure errors, whose linearized values satisfy the constraints but actual values violate;
    \item Type II: secure errors, whose actual values satisfy the constraints but linearized values violate.
\end{itemize}
$N_e$ and $err$ are the total number and the averaged value of the errors defined as follows:
\begin{subequations}
\begin{align}
    N_e &  = |\mathcal{E}|\\
    err &  = \frac{1}{N_e}\sum_\mathcal{E} \frac{ I_{F_L}^{''(\mathcal{E})}-I_{sc_F}^{''(\mathcal{E})}}{I_{sc_F}^{''(\mathcal{E})}},
\end{align}    
\end{subequations}
with $\mathcal{E}$ being the set of errors.

As a comparison, the approximation based on the Least Squares Regression (LSR) is considered first, which gives rise to the errors of both types due to their equally weighted penalty, as shown in Table~\ref{tab:SCC_unification}. Type I errors may lead to system instability whereas Type II errors may increase system operation cost due to the conservativeness. In order to eliminate the insecure operating points (Type I errors), the method in \eqref{DM3} is proposed with the performance of different complexity being evaluated in Table~\ref{tab:SCC_unification}. 

For the case of $j=0$, which corresponds to a linear model since the dimension of matrix $\mathsf{A}$ is zero, all Type I errors are eliminated with Constraint \eqref{coef_ctr2}. Moreover, the number and average value of Type II errors are also slightly reduced, as only the data points around the limit are minimized in \eqref{obj3}. Although the time to compute the coefficients is increased, it is insignificant since this process is carried out offline. Note that the case of $j = 1$ is equivalent to that of $j=0$, thus being omitted here. The SOC approximation ($j = 2$) shows a considerable improvement in terms of the number and average value of Type II errors, by introducing nonlinear elements. Furthermore, it can be observed that increasing the dimension of matrix $\mathsf{A}$ can further improve the approximation accuracy at the cost of more computational time. It should be noted that the time indicated in Table~\ref{tab:SCC_unification} is that of calculating the coefficients of the approximated expressions. It does not imply the time of solving the optimization problem after embedding the stability constraints, which is discussed at the end of Section~\ref{sec:5.2.1}.

\begin{table}[!t]
\renewcommand{\arraystretch}{1.3}
\caption{Assessment of various constraint approximation methods}
\label{tab:SCC_unification}
\noindent
\centering
    \begin{minipage}{\linewidth} 
    \renewcommand\footnoterule{\vspace*{-5pt}} 
    \begin{center}
        \begin{tabular}{ c || c | c | c | c | c  }
            \toprule
             \multirow{2}{3.3em}{\textbf{Method}} & \multicolumn{2}{c|}{\textbf{Type I Error}} &\multicolumn{2}{c|}{\textbf{Type II Error}}  & \multirow{2}{2.5em}{\textbf{Time}$\,[\mathrm{s}]$}\\ 
            \cline{2-5}
            &$N_e$ & $err$ & $N_e$ & $err$ & \\ 
            \cline{1-6} 
            LSR & $801$ & $15.55\%$ & $553$ & $-4.63\%$ & $1.6\times10^{-1}$ \\
            \cline{1-6} 
            $j=0$ & $0$ & $0$ & $404$ & $-4.31\%$ & $8.8\times10^0$  \\
            \cline{1-6} 
            $j=2$ & $0$ & $0$ & $244$ & $-2.20\%$ & $5.6\times10^1$  \\
            \cline{1-6} 
            $j=3$ & $0$ & $0$ & $203$ & $-1.55\%$ & $2.1\times10^2$ \\
           \bottomrule
        \end{tabular}
        \end{center}
    \end{minipage}
\end{table} 

\subsection{Visualization of Stability Boundaries} \label{sec:2.3}
\begin{figure}[!b]
    \centering
    \vspace{-0.4cm}
	\scalebox{1.32}{\includegraphics[trim=0 0 0 0,clip]{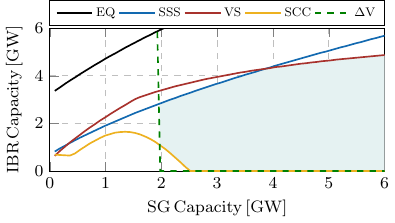}}
    \caption{\label{fig:boundaries}Boundaries of different stability constraints.}
\end{figure}
In order to visualize the boundaries of different stability constraints through a simple example, for each constraint derived in Part I of this paper, its stability boundary is evaluated in terms of the generation capacity mix for GFL IBR and SG. The results are depicted in Fig.~\ref{fig:boundaries}, where the green area represents the overall stability region. It should be noted that the impact of generator status is not considered in this section and all the generation units are assumed to be online with their capacity varying proportionally. For synchronization stability, the existence of equilibrium point (EQ) and small-signal synchronization stability (SSS) are plotted with the stable region being the lower right of the curves. It can be observed that both constraints require a certain amount of SGs to provide sufficient system strength supporting synchronization stability of GFL IBR and the existence of equilibrium points is dominated by the small-signal stability. 

Similarly, to ensure static voltage stability (VS), the maximum GFL IBR capacity cannot exceed certain limits, depending on the SG capacity. It should be noted that the static voltage stability boundary in Fig.~\ref{fig:boundaries} is evaluated without considering the reactive power support from IBR devices, which would shift the stability boundary upwards. For the transient voltage stability, the boundary to maintain a certain level of short circuit current (SCC) is represented by the yellow curve. With the feasible region being the area on the upper right of the curve, this constraint requires the system to have enough generation capacity for SCC provision. As the SG capacity decreases, the SCC from SGs cannot maintain the required SCC in the system at a certain stage (about $2.5\,\mathrm{GW}$ in this case). IBRs are thus needed to supply SCC and when the SG capacity reaches below $1.4\,\mathrm{GW}$, fewer IBRs are required since the voltage deviation at the IBR terminals becomes larger and more SCC can be provided according to the droop control strategy of the IBR. Finally, the post-fault voltage deviation constraint ($\mathrm{\Delta V}$) is represented by the green dashed curve to avoid overlapping. It defines a minimum SG capacity in the system, which is barely influenced by the IBR capacity due to its current source characteristics. 

Except for the existence of the equilibrium point of the synchronization stability, no clear dominance can be observed among other stability constraints, since the overall feasible region is defined by the intersection of all the other constraints, which highlights the importance of considering various stability constraints simultaneously. In addition, the relationship of different stability constraints would become more complicated once the actual system operating conditions and the generator status are taken into account. Therefore, it is necessary to incorporate the stability constraints into the system scheduling model, instead of dealing with them in an offline manner.

\section{Applications of Stability-Constrained Optimization} \label{sec:4}
This section introduces two applications of stability-constrained optimization in system scheduling and stability service pricing respectively. These two applications are closely related to each other. The stability-constrained system scheduling model connects the dynamics-based stability performance with the economic performance of the system during steady-state operation. It is understandable that to ensure system stability, more SGs or IBRs should be incentivized properly to provide stability services. However, the conventional duality-based pricing scheme cannot be directly applied here due to the nonconvex nature of the optimization problem. Therefore, based on the stability-constrained system scheduling, an alternate, the marginal unit price, is defined for each stability service provider.

\subsection{System Scheduling Model} \label{sec:4.1}
For a general stability-constrained optimal system scheduling problem, the objective is to minimize overall system operation cost, which is subjected to a number of constraints including power flow and power balance constraints, thermal unit constraints, and system stability constraints. 

\subsubsection{Objective Function} \label{sec:4.1.1}
The objective of the UC problem is to minimize the expected cost over all nodes in the given scenario tree:
\begin{equation}
    \label{eq:SUC}
    \min \sum_{n\in \mathcal{N}} \pi (n) \left( \sum_{g\in \mathcal{G}}  C_g(n) + \Delta t(n) c^s P^s(n) \right)
\end{equation}
where $\pi(n)$ is the probability of scenario $n\in \mathcal{N}$ and $C_g(n)$ is the operation cost of unit $g\in \mathcal{G}$ in scenario n, including startup, no-load and marginal cost; $\Delta t(n)c^sP^s(n)$ represents the cost of the load shedding in scenario n with the three terms being the time step of scenario n, load shedding cost and shed load. The scenario tree is built based on user-defined quantiles of the forecasting error distribution to capture the uncertainty associated with the demand and wind generation. \cite{conejo2010decision} can be referred to for more details.

\subsubsection{Power Balance and Power Flow Constraints}
AC power flow and its SOCP relaxation recently developed in \cite{Kocuk2016} are adapted as follows. For each line $ij\, \in \mathcal{R} $, define the ancillary variables:
\begin{subequations}
\label{Def_CS}
    \begin{align}
        c_{ij} &= |V_i||V_j|\cos(\theta_i- \theta_j) \\
        s_{ij} &= -|V_i||V_j|\sin(\theta_i- \theta_j) ,
    \end{align}
\end{subequations}
where $|V_i|\angle{\theta_i}$ is the voltage at Bus $i$. As a result, power balance and AC power flow constraints in SOC form can be formulated as: 

\begin{subequations}
\begin{align}
    & p_{t,n,i}^G = \sum_{\Omega_{g-i}} p_{t,n,g} + \sum_{\Omega_{w-i}} p_{t,n,w}  + \sum_{\Omega_{m-i}} p_{t,n,m}, \;\;\;\;\; \forall t,n,i \label{6-}\\
    & q_{t,n,i}^G = \sum_{\Omega_{g-i}} q_{t,n,g} + \sum_{\Omega_{w-i}} q_{t,n,w}  + \sum_{\Omega_{m-i}} q_{t,n,m}, \;\;\;\;\; \forall t,n,i \label{6+}\\
    & p_{t,n,i}^D = \sum_{\Omega_{l-i}} p_{t,n,l} - \sum_{\Omega_{l-i}} p_{t,n,l}^c, \;\;\;\;\; \forall t,n,i \label{7-}\\
    & q_{t,n,i}^D = \sum_{\Omega_{l-i}} q_{t,n,l} - \sum_{\Omega_{l-i}} q_{t,n,l}^c, \;\;\;\;\; \forall t,n,i \label{7+}\\
    & p_{t,n,i}^G - p_{t,n,i}^D = G_{ii}c_{ii} +  \sum_{ij\in\mathcal{R}}p_{t,n,ij}, \;\;\;\;\; \forall t,n,i\label{8-}\\
    & q_{t,n,i}^G - q_{t,n,i}^D = - B_{ii}c_{ii} + \sum_{ij\in\mathcal{R}}q_{t,n,ij}, \;\;\;\;\; \forall t,n,i\label{8+}\\
    &V_{\mathrm{min}}^2\le c_{ii} \le V_{\mathrm{max}}^2,\,\,\;\,\forall i \in \mathcal{I} \label{AC_V} \\
    & c_{ij} = c_{ji}, \,\,s_{ij} = -s_{ji} ,\,\,\;\,\forall ij \in \mathcal{R} \label{AC_cs}\\
    & c_{ij}^2+s_{ij}^2\le c_{ii}c_{jj},\,\,\;\,\forall ij \in  \mathcal{R}  \label{AC_soc}\\
    & p_{t,n,ij}^2 + q_{t,n,ij}^2 \le S_{\mathrm{max},ij}^2, \;\;\;\;\; \forall ij\in \mathcal{R}, t,n \label{10}
\end{align}
\end{subequations}
Total active power generation and load at each bus are defined in \eqref{6-} and \eqref{7-} with $\Omega_{g/w/m-i}$ and $\Omega_{l-i}$ being the set of synchronous/wind/PV units and loads connected to bus $i$. $p_{t,n,i}^{G/D}$ is the generation/demand at time $t\in\mathcal{T}$ in scenario $n\in\mathcal{N}$ at Bus $i\in\mathcal{I}$ and $p_{t,n,g/w/m}$ denotes the generated power from SG/WT/PV units. $p_{t,n,l}$ and $p_{t,n,l}^c$ are active load and load curtailment of Load $l\in\mathcal{L}$. Similarly, \eqref{6+} and \eqref{7+} define the constraints of reactive power. Power balance at each bus is given by \eqref{8-} to \eqref{8+} where $p_{t,n,ij}/q_{t,n,ij}$ is the power flow from bus $i$ to $j$ and $ij \in \mathcal{R}$ is the set of branches. $G_{ii}+\mathrm{j} B_{ii}$ is the diagonal element of nodal admittance matrix $Y$. \eqref{AC_V} is the voltage magnitude constraints at all buses with $V_{\mathrm{max}}$ and $V_{\mathrm{min}}$ being the voltage limits. \eqref{AC_cs} and \eqref{AC_soc} are the SOC relaxation of \eqref{Def_CS}. Equation \eqref{10} is the line rating constraint with $S_{\mathrm{max},ij}$ being the apparent power capacity of Line $ij$.

\subsubsection{Thermal Unit Constraints}
\begin{subequations}
    \begin{align}
        & z_{1,n,g} =x_{1,n,g}, \;\;\;\;\; \forall s,g \label{1-}\\
        & z_{t,n,g} \ge x_{t,n,g} - x_{t-1,n,g}, \;\;\;\;\; t>1,\forall s,g\\
        & x_{t,n,g} = x_{t,1,g}, \;\;\;\;\; \forall g\in \mathcal{G}_1,t,n \label{-1}\\
        &x_{t,n,g}  P_{\mathrm{min},g} \le p_{t,n,g} \le x_{t,n,g} P_{\mathrm{max},g}, \;\;\;\;\; \forall g,t,n \label{2-} \\
        &x_{t,n,g}  Q_{\mathrm{min},g} \le q_{t,n,g} \le x_{t,n,g} Q_{\mathrm{max},g}, \;\;\;\;\; \forall g,t,n \label{2+} \\
        & -r_d \le  p_{t,n,g} - p_{t-1,n,g} \le r_u , \;\;\;\;\; \forall g,t \label{ramp} \\
        &0 \le R_{t,n,g} \le x_{t,n,g} R_{\mathrm{max},g}, \;\;\;\;\; \forall g,t,n \label{R1} \\
        &0 \le R_{t,n,g} \le P_{\mathrm{max},g} - p_{t,n,g}, \;\;\;\;\; \forall g,t,n \label{R2}
    \end{align}
\end{subequations}
Equations \eqref{1-} to \eqref{-1} confine generator start-up ($z_{t,n,g}$) and on/off ($x_{t,n,g}$) statues, where $g\in \mathcal{G}_1$ is the set of inflexible generators. Power generation of thermal units, $p/q_{t,n,g}$ is bounded by their minimum and maximum limits ($P/Q_{\mathrm{max},g}$ and $P/Q_{\mathrm{min},g}$) as in \eqref{2-} and \eqref{2+}. The ramp constraint of the thermal units is considered in \eqref{ramp} with $r_d$ and $r_u$ being the ramp down and up limits. \eqref{R1} and \eqref{R2} constrain the frequency reserves $R_{t,n,g}$ from thermal units by their headroom, with $R_{\mathrm{max},g}$ being the frequency response capacity of generator $g$.

\subsubsection{System Stability Constraints} \label{sec:4.1.4}
System stability constraints can be expressed in a unified way as demonstrated in Part I of this paper:
\begin{align}
    \label{stability_ctrs}
    \Tilde{\mathbf{g}}_i  (\mathsf{X}_i) = \mathsf{c}_i \mathsf{X}_i+ \mathsf{d}_i -&\left\lVert \mathsf{A}_i \mathsf{X}_i + \mathsf{b}_i \right\lVert \ge \mathbf{g}_{i_\mathrm{lim}}, \nonumber \\
    &\qquad\qquad i\in \{1,2,3,4,5,6\}.
\end{align}

\subsection{Stability Service Pricing}
In order to provide economic incentives for generation units, according to the value they create for stabilizing the system, a proper stability market mechanism should be established. However, the conventional pricing scheme based on marginal shadow price in the energy market cannot be directly applied to stability market due to the following reasons. It is not clear which services should be defined in the stability market as generation units may provide multiple services for different stability enhancement. \textcolor{black}{Whether these services such as inertia provision and system strength enhancement, can be provided to the system depends on the commitment decision of SGs. Hence, these services vary in a discontinuous (binary) manner and cannot be separated from each other.} Moreover, the stability services provided by SGs and GFM IBRs interact with each other due to the strong coupling between different stability constraints. Additionally, the stability services of SGs depend on their commitment decisions, which brings in binary variables and nonconvexity, making the marginal price being ill-defined.

As a result, the concept of marginal unit price proposed in \cite{MUKHERJEE200653} is adapted in the system scheduling model to quantify the economic value of stability enhancement provided by different generation units, where a single price is defined for each SG or GFM IBR according to its contribution to system stability.  Denote the optimization problem defined in Section~\ref{sec:4.1} in a compact way as:
\begin{subequations}
\label{UC_compact}
    \begin{align}
        \min_{\mathsf{X}}\;\; &\mathsf{f} (\mathsf{X})\\
        \mathrm{s.t.}\;\; & \Tilde{\mathbf{g}}_i  (\mathsf{X}) = \mathsf{c}_i \mathsf{X}+ \mathsf{d}_i - \left\lVert \mathsf{A}_i \mathsf{X} + \mathsf{b}_i \right\lVert \ge \mathbf{g}_{i_\mathrm{{lim}}},\quad \forall i\\
        & \mathbf{k}(\mathsf{X})\ge 0,
    \end{align}
\end{subequations}
where $\mathsf f(\mathsf{X})$ is system operation cost, $\Tilde{\mathbf{g}}_i  (\mathsf{X})$ are system stability constraints and $\mathbf{k}(\mathsf{X})$ is the rest of the constraints. Denote the optimal value of problem \eqref{UC_compact} as $\mathsf{f}^*$. To determine the economic value of a generator (corresponding to the $g^\mathrm{th}$ element in $\mathsf{X}$) on system stability improvement, the following problem is solved.
\begin{subequations}
\label{UC_delta}
    \begin{align}
        \min_{\mathsf{X}}\;\; &\mathsf{f} (\mathsf{X}) \label{f_X2}\\
        \mathrm{s.t.}\;\; & \Delta \Tilde{\mathbf{g}}_i  (\mathsf{X}) = (\mathsf{c}_i+\Delta \mathsf{c}_i^g) \mathsf{X}+ \mathsf{d}_i \nonumber \\
        & \qquad\quad\; - \left\lVert \left(\mathsf{A}_i+\Delta \mathsf{A}_i^g\right) \mathsf{X} + \mathsf{b}_i \right\lVert \ge \mathbf{g}_{i_\mathrm{{lim}}}\quad ,\forall i \label{Delta_g}\\
        & \mathbf{k}(\mathsf{X})\ge 0 \label{k_X2}.
    \end{align}
\end{subequations}
The matrices $\Delta \mathsf{c}_i^g$ and $\Delta \mathsf{A}_i^g$ are of the same dimension as $\mathsf{c}_i$ and $\mathsf{A}_i$, which can be expressed as:
\begin{subequations}
\label{Delta_cA}
    \begin{align}
    \Delta \mathsf{c}_{i,k}^g =  &
    \begin{cases}
        0\qquad\quad\;\;\, ,\mathrm{if}\, k\neq g\\
        -\mathsf{c}_{i,k}\qquad,\mathrm{if}\,k = g\\
    \end{cases}\\
    \Delta \mathsf{A}_{i,k}^g = &
    \begin{cases}
        0\qquad\quad\;\;\,\, ,\mathrm{if}\, k\neq g\\
        -\mathsf{A}_{i,k}\qquad,\mathrm{if}\,k = g\\
    \end{cases}
    \end{align}
\end{subequations}
where $(\cdot)_{,k}$ denotes the $k^\mathrm{th}$ column of the matrix. With the above definition, the coefficients in $\mathsf{c}_i+\Delta \mathsf{c}_i^g$ corresponding to the $g^\mathrm{th}$ element in $\mathsf{X}$, become zero, thus eliminating the impact of unit $g$ on system stability, whilst its impacts on the operation cost \eqref{f_X2} and other constraints \eqref{k_X2} remain the same. Denoting the optimal value of problem $\eqref{UC_delta}$ as $\mathsf{f}^{*}_g$ enables us to define the economic value of unit $g$ on stabilizing the system:
\begin{equation}
\label{mu_g}
    \mu_g = \mathsf{f}^{*}_g-\mathsf{f}^{*},
\end{equation}
with $\mu_g$ being the price of stability service provided by generator $g$.

\section{Results} \label{sec:5}
In order to investigate the system stability at different wind penetrations, and how stability constraints influence system operation, we consider an optimization problem with a time horizon of 24 hours and a time step of 1 hour, which is solved by Gurobi (10.0.0) on a PC with Intel(R) Core(TM) i7-7820X CPU @ 3.60GHz and RAM of 64 GB. The weather conditions are obtained from online numerical weather prediction \cite{weather}. The frequency limits of nadir, steady-state value and RoCoF are set as: $0.8\,\mathrm{Hz}$, $0.5\,\mathrm{Hz}$ and $0.5\,\mathrm{Hz/s}$ respectively. The critical gSCR is $\mathrm{gSCR}_{\mathrm{lim}}=2.5$. The limits for post-fault short circuit current and transient voltage dip are $20\,\mathrm{p.u.}$ and $0.15\,\mathrm{p.u.}$ respectively.

\subsection{Stability Assessment}
\begin{figure}[!t]
    \centering
	\scalebox{1.25}{\includegraphics[trim=0 0 0 0,clip]{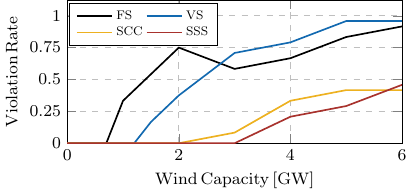}}
    \caption{\label{fig:V_Rate}Stability violation rate at different wind penetration.}
\end{figure}

In this subsection, a traditional steady-state only system scheduling model, without any stability constraints discussed in Section~\ref{sec:4.1.4}, is performed and the resulting system operation conditions are used to assess the system stability as wind penetration increases. The violation rate during 24 hours is shown in Fig.~\ref{fig:V_Rate}, with four stability constraints being considered, namely frequency stability (FS), small-signal/static voltage stability (VS), short-circuit current (SCC), and small-signal synchronization stability (SSS). The rest two stability constraints are either dominated by other stability constraints as illustrated in Fig.~\ref{fig:boundaries} (EQ) or not violated during the entire time horizon ($\mathrm{\Delta V}$), hence not being covered in the figure.

For FS, constraint violation starts to appear when the system wind capacity approaches about $1\,\mathrm{GW}$ and increases to around $75\%$ at $2\,\mathrm{GW}$ wind penetration, which is due to the declined frequency response and inertia from conventional SGs. A violation rate reduction is noticed between $2\,\mathrm{GW}$ and $3\,\mathrm{GW}$ wind penetration as the increased wind power is not large enough to have SGs be disconnected and the reduced power output from SGs leads to larger headroom for providing frequency response. However, after the wind capacity continues increasing, the violation rate of frequency stability rises again to above $90\%$ at $6\,\mathrm{GW}$ wind penetration. 

Regarding VS, a continuously increasing trend is observed between $1.2\,\mathrm{GW}$ and $5\,\mathrm{GW}$ wind penetration, because of the increased power injection and declined system strength at the IBR buses. The increment rate gradually reduces except at $4\,\mathrm{GW}$ wind penetration and becomes 0 after $5\,\mathrm{GW}$ wind penetration. Moreover, the violation rate of VS exceeds that of FS before the wind capacity reaches to $3\,\mathrm{GW}$.

The violation rates for the SCC and SSS constraints appear at higher wind penetration and are relatively smaller compared with the previous two. They start to increase at about $2\,\mathrm{GW}$ and $3\,\mathrm{GW}$ wind penetration respectively and reach to more than $40\%$ at $6\,\mathrm{GW}$ wind penetration. In addition, the varying increment rates of all four stability constraints also indicate the nonlinear and intricate impact of the wind capacity on system stability. Nevertheless, it is clearly demonstrated that system stability constraints are violated to different extents as the wind penetration increases, if no stability constraints are embedded in the system scheduling model. 

\subsection{Impact of Stability Constraints on System Scheduling} \label{sec:5.2}
\subsubsection{System operation cost} \label{sec:5.2.1}
In order to maintain system stability, the stability constraints are incorporated into the system scheduling model and their interactions and impacts on system operation are investigated. Table~\ref{tab:Cost_Stability} gives the system operation cost increment compared with the case where no stability constraints are included. All 16 possible combinations of the four stability constraints in Fig.~\ref{fig:V_Rate} are considered and the wind capacity is set as $5\,\mathrm{GW}$. Moreover, to illustrate the impact of Synthetic Inertia (SI) provision, another dimension (SI=0/1) is added to the table. For ease of comparison, the numbers in the table are collected in matrix $\mathsf{C}$, with $\mathsf{C}_{ij}$ being the element of $i^\mathrm{th}$ row and $j^\mathrm{th}$ column in the table. As the baseline, the averaged operational cost among the concerned time horizon, corresponding to $\mathsf{C}_{11}$ where no stability constraints are considered, is $236.75\,\mathrm{k\pounds/h}$.

\begin{table}[!t]
\vspace{-0.4cm}
\renewcommand{\arraystretch}{1.3}
\caption{System operation cost increment due to different stability constraints}
\label{tab:Cost_Stability}
\noindent
\centering
    \begin{minipage}{\linewidth} 
    \renewcommand\footnoterule{\vspace*{-5pt}} 
    \begin{center}
        \begin{tabular}{ c | c || c | c | c | c }
            \toprule
             \multicolumn{2}{c||}{\textbf{Cost Increment}} & \multicolumn{2}{c|}{\textbf{SCC = 0}} &\multicolumn{2}{c}{\textbf{SCC = 1}} \\ 
             \cline{3-6} 
             \multicolumn{2}{c||}{$\mathbf{[k\pounds/h]}$} &\textbf{SSS = 0} & \textbf{SSS = 1} & \textbf{SSS = 0} & \textbf{SSS = 1} \\ 
            \cline{1-6} 
            \multirow{2}{*}{\textbf{FS = 0}} & \textbf{VS = 0} & \gradient{0} & \gradient{15.36} & \gradient{30.23} & \gradient{50.45}  \\
            \cline{2-6} 
             & \textbf{VS = 1} & \gradient{14.87} & \gradient{17.65} & \gradient{38.13} & \gradient{54.08}   \\
            \cline{1-6} 
            {\textbf{FS = 1}} & \textbf{VS = 0} & \gradient{74.56} & \gradient{76.17} & \gradient{75.67} & \gradient{77.34}\\
            \cline{2-6} 
            {\textbf{(SI = 0)}} & \textbf{VS = 1} & \gradient{75.87} & \gradient{77.42} & \gradient{76.89} & \gradient{77.82}\\
             \cline{1-6} 
           {\textbf{FS = 1}} & \textbf{VS = 0} & \gradient{13.63} & \gradient{24.32} & \gradient{30.62} & \gradient{51.87}\\
            \cline{2-6} 
           {\textbf{(SI = 1)}}  & \textbf{VS = 1} & \gradient{24.70} & \gradient{26.22} & \gradient{38.62} & \gradient{54.31}\\
           \bottomrule
        \end{tabular}
        \end{center}
    \end{minipage}
\end{table} 

First, it is noticed that if no SI is provided by IBRs, the system stability is dominated by FS. To maintain frequency stability, additional SGs have to be committed online to provide inertia and frequency responses, leading to significant cost increment ($\mathsf{C}_{3j}$ and $\mathsf{C}_{4j}$). With these SGs, other stability constraints can be automatically maintained during most of the time. Hence, whether or not considering other stability constraints makes few differences in terms of system operation cost. Once SI is available from IBRs, system operation cost can be reduced significantly as in $\mathsf{C}_{5j}$ and $\mathsf{C}_{6j}$, which are further analyzed together with $\mathsf{C}_{1j}$ and $\mathsf{C}_{2j}$ to investigate the interaction of different stability constraints. 

It is found that, in general, incorporating more stability constraints (moving towards the bottom-right corner of the table) leads to more system operation cost where an increment of $23\%$ is observed at $\mathsf{C}_{64}$. Furthermore, the impact of different stability constraints on the system scheduling varies. For FS, a cost increment of $13.63\,\mathrm{k\pounds/h}$ is calculated by taking the difference between $\mathsf{C}_{51}-\mathsf{C}_{11}$. Similarly, the averaged cost increments for VS, SCC and SSS are $14.87$, $30.23$ and $15.36\,\mathrm{k\pounds/h}$ respectively.


Additionally, different stability constraints interact with one another. For the frequency stability, it can be deduced that incorporating SCC constraints helps to maintain FS as well, by comparing $\mathsf{C}_{31}-\mathsf{C}_{11}$ ($13.63\,\mathrm{k\pounds/h}$, averaged cost to maintain FS without SCC constraints) with $\mathsf{C}_{33}-\mathsf{C}_{31}$ ($0.40\,\mathrm{k\pounds/h}$, averaged cost to maintain FS with SCC constraints). This is because 
maintaining SCC requires more SGs in the system thus facilitating the frequency stability. Likewise, the incorporation of VS and SSS constraints also reduces the cost to maintain FS from $13.63\,\mathrm{k\pounds/h}$ to $9.84\,\mathrm{k\pounds/h}$ and $8.96\,\mathrm{k\pounds/h}$ respectively, but the effect is less significant. 

With the same reasoning, similar observations can be made: i) VS can be dramatically improved by the SSS constraints with the VS maintenance cost being reduced from $14.87\,\mathrm{k\pounds/h}$ to $2.29\,\mathrm{k\pounds/h}$ but the influences due to FS and SCC constraints are less obvious; ii) SCC constraints becomes easier to maintain with the frequency constraints included, but are less impacted by VS and SSS; iii) SSS can be maintained with lower additional cost after including VS constraint whereas the cost reduction after including FS and SCC constraints is not considerable. In summary, FS and SCC constraints relate more with each other whilst the connection and interaction between VS and SSS are closer and stronger. 

\subsubsection{Computational Performance} 
The computational time of the proposed approach is assessed against alternative assumptions. In particular, three cases are compared as shown in Table~\ref{tab:time} where different numbers of scenarios ($|\mathcal{N}|$) are considered. In the case where the linearized AC power flow \cite{trodden2013optimization} is applied and no stability constraints are included, the computational time is the shortest, since the optimization only contains linear constraints leading to an MILP formulation. After incorporating the SOC-based AC power flow relaxation, the computational time increases considerably due to the more complex formulation of MISOCP. On top of that, we further add the stability constraints in the scheduling model which is represented by the case of ``SOC AC + SC'' in the table. It can be observed that although more time is required to solve the model, the time increment compared with the ``SOC AC'' case is not significant, indicating that the additional constraints in the proposed stability-constrained optimization do not present computational issues. Moreover, as more scenarios in the stochastic optimization are considered, the computation time increases especially for the MISOCP formulation, because of the complex interaction of different scenarios and the SOC constraints. Nevertheless, it is clear that the proposed stability-constrained optimization does not increase the computation time significantly compared with the case where SOC-based AC power flow formulation is used.

\begin{table}[!t]
\vspace{-0.4cm}
\renewcommand{\arraystretch}{1.2}
\caption{Comparison of Computational Time}
\label{tab:time}
\noindent
\centering
    \begin{minipage}{\linewidth} 
    \renewcommand\footnoterule{\vspace*{-5pt}} 
    \begin{center}
        \begin{tabular}{ c || c | c | c }
            \toprule
             Time & Linearized AC & SOC AC & SOC AC + SC \\ 
             $\mathrm{[s/step]}$ & (MILP) & (MISOCP) & (MISOCP)\\
            \cline{1-4}
            $|\mathcal{N}|=1$& 0.23 & 0.32 & 0.46\\
            $|\mathcal{N}|=7$& 0.31 & 3.92 & 4.07\\
            $|\mathcal{N}|=15$& 0.48 & 124.44 & 135.79\\
           \bottomrule
        \end{tabular}
    \end{center}
    \end{minipage}
\end{table} 

\subsubsection{Synchronous generator status}
\begin{figure}[!b]
    \centering
    \vspace{-0.4cm}
	\scalebox{1.25}{\includegraphics[trim=0 0 0 0,clip]{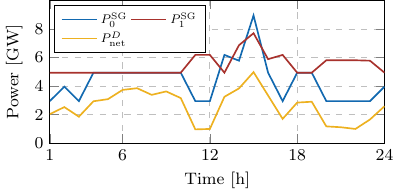}}
    \caption{\label{fig:Num_SG}Unit commitment results and system net demand during 24 hours.}
\end{figure}

The operation cost investigated in Section~\ref{sec:5.2.1} closely relates to the commitment states of conventional generators in the system. Therefore, to better reveal how the stability constraints impact the SG commitment decisions, the total online capacity of SG with and without stability constraints ($P_1^{\mathrm{SG}}$ and $P_0^{\mathrm{SG}}$) is plotted in Fig.~\ref{fig:Num_SG}, together with system net demand $P^D_{\mathrm{net}}$. 

It is understandable that the committed SG capacity in the case where no stability constraints (blue curve) are incorporated follows the net demand (yellow curve) in the system, in order to ensure the power balance. However, once the stability constraints are included in the scheduling model (red curve), the committed SG capacity increases by about $3\,\mathrm{GW}$ when the net demand is low, i.e., $11-12$, $16-18$ and $20-23\,\mathrm{h}$. This is due to the fact that during these hours, the system becomes weak with the main source in the system being GFL units, therefore additional SGs are needed to maintain different stability. For the rest of the time where the system net demand is higher, the online SG capacities are comparable in the two cases.  

For those SGs that are committed for the sole purpose of stability maintenance, they may not be able to recover their operation cost by only participating in the energy market due to high renewable generation and relatively low clearing price. Hence, it is necessary to design a stability market where appropriate incentives can be provided for conventional SGs and GFM IBRs to enhance system stability. More details regarding the stability market design are discussed in Section~\ref{sec:5.4}.

\subsubsection{Assessment of Stability Margin}
\begin{figure}[!t]
    \centering
    \vspace{-0.4cm}
	\scalebox{1.25}{\includegraphics[trim=0 0 0 0,clip]{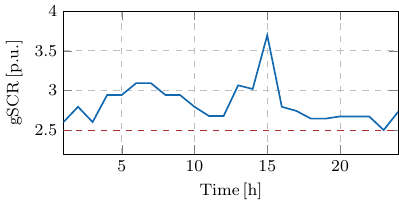}}
    \caption{\label{fig:Margin1}System gSCR and its limit during one-day operation.}
    \vspace{-0.4cm}
\end{figure}

The stability margin is also assessed here, where only the example of small-signal synchronization stability is taken as other stability indices present a similar trend. The results of the actual $\mathrm{gSCR}$ and its limit during one-day operation are illustrated in Fig.~\ref{fig:Margin1}. It can be observed that during some of the hours, the actual gSCR is much higher than the limits ($\mathrm{[3-11], \,[12-17]\,h}$). On one hand, this is due to the discrete variation of the SGs' status. On the other hand, the gSCR constraint can be nonbinding when it is dominated by other stability constraints during some hours. Noticeably, there is a large value of gSCR at the $15^{\mathrm{th}}$ hour because of the high net demand as shown in Fig.~\ref{fig:Num_SG}. During the rest of the time, the gSCR is close to the limit, indicating an acceptable conservativeness.

\subsection{Impact of synthetic inertia and reactive power provision from IBRs}
\begin{figure}[!b]
    \centering
    \vspace{-0.4cm}
	\scalebox{1.25}{\includegraphics[trim=0 0 0 0,clip]{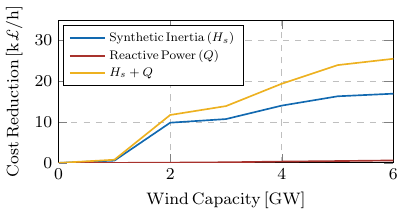}}
    \caption{\label{fig:Services}System operation cost reduction due to voltage and frequency support.}
\end{figure}
The value of synthetic inertia has been partly revealed in Table~\ref{tab:Cost_Stability}. In order to further investigate the impact of synthetic inertia and reactive power provision from IBRs on system operation at different wind penetrations, simulations of four different cases are performed with all the stability constraints being embedded. The results are illustrated in Fig.~\ref{fig:Services} by computing the cost reduction compared with the case where neither synthetic inertia nor reactive power is provided. It can be observed that the economic value of synthetic inertia increases with wind capacity. This is because on one hand, the frequency stability issue becomes severer as wind capacity rises and on the other hand, the increased wind capacity results in more capability to provide synthetic inertia. This trend continues until a certain wind capacity (around $5\, \mathrm{GW}$ in this case), after which the amount of synthetic inertia becomes more than enough and its economic value saturates. 

For the reactive power support, its economic value is not obvious if no synthetic inertia is provided by IBRs, since in this case the static voltage stability can be maintained by the SGs required by frequency constraints, which is consistent with the findings obtained from Table~\ref{tab:Cost_Stability}. Once synthetic inertia becomes available, the static voltage stability would also become an issue, thus increasing the economic value of reactive power support, which is indicated by the difference between yellow and blue curves. Nevertheless, the effectiveness and economic value of frequency and voltage support from IBR devices in maintaining frequency and voltage stability can be clearly demonstrated.

\subsection{Implications for stability market design}\label{sec:5.4}
Simulation results related to the stability service prices are reported in this section with the corresponding implication for stability market design being discussed.

\subsubsection{Stability service prices during 24-hour system operation}
\begin{figure}[!t]
    \centering
    \vspace{-0.4cm}
	\scalebox{1.25}{\includegraphics[trim=0 0 0 0,clip]{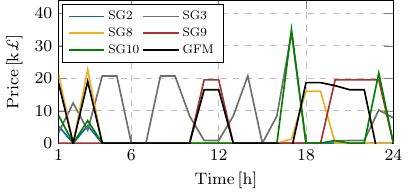}}
    \caption{\label{fig:Price_time}Stability service price of generation units within 24-hour dispatch.}
    \vspace{-0.4cm}
\end{figure}
The stability service price defined in \eqref{mu_g} is evaluated for each SG and GFM IBR within an operation period of 24 hours, with the installed wind capacity being $5\,\mathrm{GW}$. The results are depicted in Fig.~\ref{fig:Price_time}, where five SGs and one GFM IBR have non-zero prices during the day. It can be observed that the stability service prices for various units are different which are proportional to their contribution to system stability. In general, the units that are close to GFL IBRs and have larger capacities tend to contribute more to system stability. The prices also vary significantly during the 24 hours, which is impacted by the net demand in the system. During high net demand periods, i.e., $6-7$ and $15\,\mathrm{h}$, the prices of all units are zero, since the SGs required to meet the demand can ensure system stability leading to unbinding stability constraints. 

The price of GFM IBR is also given in the figure (black curve), which is comparable to the value of other SGs. This price presents a resemblant pattern to that of SG8 during $1-4$, $17-19\,\mathrm{h}$ and SG9 during $10-13$, $20-22\,\mathrm{h}$, indicating the effectiveness of GFM IBR on system stability improvement. It can be explained by the fact that these three units are electrically close to each other and the GFM IBR has a similar yet smaller electrical distance to the GFL IBRs compared with SG8 and SG9. Hence, their contributions to the system stability issues caused by the GFL IBRs are also similar.

\subsubsection{Impact of wind penetration}
\begin{figure}[!t]
    \centering
    \vspace{-0.4cm}
	\scalebox{1.12}{\includegraphics[trim=0 0 0 0,clip]{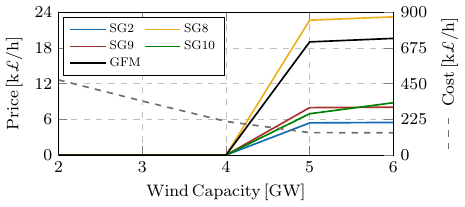}}
    \caption{\label{fig:Price_wind}Stability service price of   generation units with different wind capacity.}
    \vspace{-0.4cm}
\end{figure}
It is understandable that the value of stability services is influenced by the level of system wind penetration. In order to further investigate this relationship, the stability service prices of different units at $3\,\mathrm{h}$ are calculated with different wind capacities. As shown in Fig.~\ref{fig:Price_wind},
the stability service prices for all the units at this hour are zero until the wind capacity in the system reaches $4\,\mathrm{GW}$, meaning the stability constraints are not effective. As the wind capacity continues increasing, the stability service prices of different units rise and become saturated after about $5\,\mathrm{GW}$ wind capacity. This can be justified by the trend of system operation cost at this hour as indicated by the grey dashed curve, which ceases to decrease after $5\,\mathrm{GW}$ wind capacity. The system operating conditions are barely impacted even if more wind capacity is installed due to the stability constraints. Therefore, stability service prices of different units almost remain the same. In this case, the system operators should install more GFM IBRs, such that more renewable energy can be utilized while maintaining system stability. However, the optimal amount and position of GFM IBRs that can achieve both system stability and maximum economic benefit is beyond the scope of this work.

\section{Conclusion} \label{sec:6}
This paper validates the system stability constraints that are developed in Part I and demonstrates good accuracy and conservativeness of the unification process. Moreover, these system-level stability constraints are further applied to the system scheduling model leading to an overall MISOCP-based UC problem. Optimal system operation can be achieved with the proposed model while maintaining system stability. Thorough case studies illustrate the impacts of different stability constraints on system operation and the interactions among those constraints. The effects of system frequency and voltage support on improving system stability are presented as well. Moreover, it is also revealed that system stability is maintained by SGs, and GFM IBRs, of which the stability service value is quantified through the proposed marginal unit pricing scheme.

\bibliographystyle{IEEEtran}
\bibliography{bibliography}
\end{document}